\documentclass[aps,prc,twocolumn,superscriptaddress,showpacs]{revtex4}
\usepackage{graphicx}
\usepackage{rotating}
\usepackage{color}

\begin{document}

%\preprint{}

\title{Structure of {\boldmath $^{55}$}Sc and development of the {\boldmath $N=34$} subshell closure}


\author{D.~Steppenbeck}
\email[]{steppenbeck@riken.jp}
%\thanks{}
\affiliation{RIKEN Nishina Center, 2-1, Hirosawa, Wako, Saitama 351-0198, Japan}
\author{S.~Takeuchi}
\affiliation{Department of Physics, Tokyo Institute of Technology, Meguro, Tokyo 152-8551, Japan}
\author{N.~Aoi}
\affiliation{Research Center for Nuclear Physics, University of Osaka, Ibaraki, Osaka 567-0047, Japan}
\author{P.~Doornenbal}
\affiliation{RIKEN Nishina Center, 2-1, Hirosawa, Wako, Saitama 351-0198, Japan}
\author{M.~Matsushita}
\affiliation{Center for Nuclear Study, University of Tokyo, Hongo, Bunkyo, Tokyo 113-0033, Japan}
\author{H.~Wang}
\affiliation{RIKEN Nishina Center, 2-1, Hirosawa, Wako, Saitama 351-0198, Japan}
\author{H.~Baba}
\affiliation{RIKEN Nishina Center, 2-1, Hirosawa, Wako, Saitama 351-0198, Japan}
\author{S.~Go}
\altaffiliation[Present address:~]{RIKEN Nishina Center, 2-1, Hirosawa, Wako, Saitama 351-0198, Japan.}
\affiliation{Center for Nuclear Study, University of Tokyo, Hongo, Bunkyo, Tokyo 113-0033, Japan}
\author{J.~D.~Holt}
\affiliation{TRIUMF, 4004 Wesbrook Mall, Vancouver, British Columbia V6T 2A3, Canada}
\author{J.~Lee}
\altaffiliation[Present address:~]{Department of Physics, University of Hong Kong, Pokfulam Road, Hong Kong.}
\affiliation{RIKEN Nishina Center, 2-1, Hirosawa, Wako, Saitama 351-0198, Japan}
\author{K.~Matsui}
\affiliation{Department of Physics, University of Tokyo, Hongo, Bunkyo, Tokyo 113-0033, Japan}
\author{S.~Michimasa}
\affiliation{Center for Nuclear Study, University of Tokyo, Hongo, Bunkyo, Tokyo 113-0033, Japan}
\author{T.~Motobayashi}
\affiliation{RIKEN Nishina Center, 2-1, Hirosawa, Wako, Saitama 351-0198, Japan}
\author{D.~Nishimura}
\affiliation{Department of Physics, Tokyo City University, Setagaya, Tokyo 158-8557, Japan}
\author{T.~Otsuka}
\altaffiliation[Present address:~]{RIKEN Nishina Center, 2-1, Hirosawa, Wako, Saitama 351-0198, Japan.}
\affiliation{Center for Nuclear Study, University of Tokyo, Hongo, Bunkyo, Tokyo 113-0033, Japan}
\affiliation{Department of Physics, University of Tokyo, Hongo, Bunkyo, Tokyo 113-0033, Japan}
\author{H.~Sakurai}
\affiliation{RIKEN Nishina Center, 2-1, Hirosawa, Wako, Saitama 351-0198, Japan}
\affiliation{Department of Physics, University of Tokyo, Hongo, Bunkyo, Tokyo 113-0033, Japan}
\author{Y.~Shiga}
\affiliation{Department of Physics, Rikkyo University, Toshima, Tokyo 171-8501, Japan}
\author{P.-A.~S{\"o}derstr{\"o}m}
\altaffiliation[Present address:~]{Institut f{\"u}r Kernphysik, Technische Universit{\"a}t Darmstadt, 64289 Darmstadt, Germany.}
\affiliation{RIKEN Nishina Center, 2-1, Hirosawa, Wako, Saitama 351-0198, Japan}
\author{S.~R.~Stroberg}
\affiliation{TRIUMF, 4004 Wesbrook Mall, Vancouver, British Columbia V6T 2A3, Canada}
\author{T.~Sumikama}
\altaffiliation[Present address:~]{RIKEN Nishina Center, 2-1, Hirosawa, Wako, Saitama 351-0198, Japan.}
\affiliation{Department of Physics, Tohoku University, Sendai, Miyagi 980-8578, Japan}
\author{R.~Taniuchi}
\affiliation{RIKEN Nishina Center, 2-1, Hirosawa, Wako, Saitama 351-0198, Japan}
\affiliation{Department of Physics, University of Tokyo, Hongo, Bunkyo, Tokyo 113-0033, Japan}
\author{J.~A.~Tostevin}
\affiliation{Department of Physics, University of Surrey, Guildford, Surrey GU2 7XH, United Kingdom}
\author{Y.~Utsuno}
\affiliation{Japan Atomic Energy Agency, Tokai, Ibaraki 319-1195, Japan}
\author{J.~J.~Valiente-Dob{\'o}n}
\affiliation{Istituto Nazionale di Fisica Nucleare, Laboratori Nazionali di Legnaro, Legnaro 35020, Italy}
\author{K.~Yoneda}
\affiliation{RIKEN Nishina Center, 2-1, Hirosawa, Wako, Saitama 351-0198, Japan}


\date{\today}

\begin{abstract}
  The low-lying structure of $^{55}$Sc has been investigated using in-beam $\gamma$-ray spectroscopy with
  the $^{9}$Be($^{56}$Ti,$^{55}$Sc+$\gamma$)$X$ one-proton removal and $^{9}$Be($^{55}$Sc,$^{55}$Sc+$\gamma$)$X$
  inelastic-scattering reactions at the RIKEN Radioactive Isotope Beam Factory. Transitions with energies of
  572(4), 695(5), 1539(10), 1730(20), 1854(27), 2091(19), 2452(26), and 3241(39) keV are reported, and a level
  scheme has been constructed using $\gamma\gamma$ coincidence relationships and $\gamma$-ray relative intensities.
  The results are compared to large-scale shell-model calculations in the $sd$-$pf$ model space, which account for
  positive-parity states from proton-hole cross-shell excitations, and to {\it ab initio} shell-model calculations
  from the in-medium similarity renormalization group that includes three-nucleon forces explicitly. The results
  of proton-removal reaction theory with the eikonal model approach were adopted to aid identification of
  positive-parity states in the level scheme; experimental counterparts of theoretical $1/2^{+}_{1}$ and
  $3/2^{+}_{1}$ states are suggested from measured decay patterns. The energy of the first $3/2^{-}$ state,
  which is sensitive to the neutron shell gap at the Fermi surface, was determined. The result indicates
  a rapid weakening of the $N=34$ subshell closure in $pf$-shell nuclei at $Z>20$, even when only a single
  proton occupies the $\pi f_{7/2}$ orbital.
\end{abstract}

%\keywords{}

\maketitle

\section{INTRODUCTION}
Investigations of exotic, radioactive nuclei---isotopes that lie far from the line of $\beta$ stability on the chart
of nuclides---have highlighted structural changes that occur relative to stable systems \cite{gad08-2,sor08} owing
to differences in the ordering of single-particle orbitals that define the traditional nuclear shell model \cite{may49,hax49}.
A few noteworthy examples of such phenomena include the onset of a neutron shell gap at $N=16$ along the oxygen isotopic
chain \cite{oza00,hof08,hof09,kan09}, and the weakening of the traditional neutron magic numbers $N=20$ and $28$ in nuclei
around $^{32}$Mg \cite{gui84,mot95} and $^{42}$Si \cite{bas07,tak12}, respectively. In the neutron-rich $pf$ shell, which
is bounded by the proton and neutron numbers $Z=20$--$28$ and $N=28$--$40$, the onset of new subshell closures at $N=32$
and $34$ have received much attention on both the experimental and theoretical fronts. Development of the $N=32$ subshell
gap was first suggested from a decay study of $^{52}$K by Huck {\it et al.}~\cite{huc85}, and confirmed more recently
along the Ca \cite{gad06,wie13}, Ti \cite{jan02,din05}, and Cr \cite{pri01,bur05,zhu06} isotopic chains from investigations
of first $2^{+}$ state energies [$E(2^{+}_{1})$], reduced transition probabilities [$B(E2; 0^{+}_{1}\rightarrow 2^{+}_{1})$],
and high-precision mass measurements. The first direct evidence for the onset of a new subshell closure at $N=34$ in exotic
Ca isotopes was presented from the structure of $^{54}$Ca \cite{ste13-1}, while earlier studies on $^{56}$Ti
\cite{din05,lid04-1} indicated that no significant $N=34$ subshell closure resides in titanium isotopes. Moreover,
the persistence of the $N=32$ subshell gap below the $Z=20$ shell closure has been reported in exotic K \cite{ros15}
and Ar \cite{ste15-1} isotopes; however, recent evidence indicating a large, unexpected increase in the nuclear charge
radii of neutron-rich Ca isotopes beyond $N=28$ has emerged from laser spectroscopy experiments \cite{gar16}, which may
challenge the proposition of a significant $N=32$ subshell closure. On the theoretical side, the developments of $N=32$
and $34$ subshell gaps have been investigated, for example, in the framework of tensor-force-driven shell evolution
\cite{ots05,ots13-1}, which indicates that a weakening of the attractive proton-neutron ($\pi$-$\nu$) interaction
between the $\pi f_{7/2}$ and $\nu f_{5/2}$ orbitals in isotones approaching $Z=20$ is responsible for the appearance
of these closures in exotic systems. Much effort has also been afforded to theoretical calculations that employ
three-nucleon forces (3NFs) \cite{heb15}; some examples along the oxygen and calcium isotopic chains include
investigations of nuclear masses \cite{gal12,wie13,her14,som14}, charge radii \cite{eks15,gar16,lap16}, energy
systematics \cite{hol12,hag12,ots13-2,hol13,hol14,jan14}, electromagnetic moments \cite{gar15}, the location of
the neutron drip line \cite{ots10,ots13-1}, and very recently, the neutron distribution and skin thickness in the
doubly magic nucleus $^{48}$Ca \cite{hag16-1}, and the impact on spectroscopic factors \cite{cra17}. Theoretical
interactions involving 3NFs have also been applied to investigate the structure of the medium-mass nucleus $^{78}$Ni
\cite{hag16-2}. Furthermore, advances in many-body methods now allow for the construction of shell-model Hamiltonians
in a fully {\it ab initio} manner \cite{bog14,jan14,dik15}. In particular, when the valence-space formulation of the
in-medium similarity renormalization group (VS-IM-SRG) \cite{tsu12,str16,her16} is combined with the ensemble normal
ordering procedure introduced in Ref.~\cite{str17}, {\it ab initio} calculations can be extended to ground and excited
states of essentially all light- and medium-mass nuclei with an accuracy comparable to that in closed-shell systems.
Here, the VS-IM-SRG has been used to perform the first calculations of spectroscopy in neutron-rich Sc isotopes.

The structure of neutron-rich Sc isotopes around $N=34$ has been investigated over recent years using
$\beta$ decay, multinucleon transfer, and nucleon-knockout reactions. For the $N=32$ isotope, $^{53}$Sc,
$\gamma$-ray transitions have been reported to depopulate states at 2283(18) and 2617(20) keV from multinucleon
transfer with the $^{238}$U+$^{48}$Ca reaction in inverse kinematics \cite{bha09}, and a single transition at
2109.0(3) keV was deduced from the $\beta$ decay of $^{53}$Ca \cite{cra09,cra10}, which was placed in the level
scheme feeding the ground state directly. These excited states were assigned tentative spin-parity quantum numbers
of $9/2^{-}$, $11/2^{-}$, and $3/2^{-}$, respectively. In Refs.~\cite{cra09,cra10}, the structure of $^{53}$Sc was
discussed in the context of the extreme single-particle model; the coupling of the valence $\pi f_{7/2}$ proton
to the first excited $2^{+}$ state of $^{52}$Ca [$\pi f_{7/2}$$\otimes$$^{52}$Ca$(2^{+}_{1})$] is expected to
produce a quintet of states with spins and parities of $3/2^{-}$, $5/2^{-}$, $7/2^{-}$, $9/2^{-}$, and $11/2^{-}$.
The fact that the three excited states in $^{53}$Sc, reported at 2.11, 2.28, and 2.62 MeV \cite{bha09,cra09,cra10},
lie at energies comparable to that of the $2^{+}_{1}$ state in $^{52}$Ca (2.56 MeV \cite{huc85,gad06}), and were
assigned spin-parity values consistent with the members of the expected quintet, highlights the success of the
simple coupling scheme in this particular case and provides support for a robust $N=32$ subshell gap. The 2.11-MeV
state was later confirmed using the $^{9}$Be($^{54}$Ti,$^{53}$Sc+$\gamma$)$X$ one-proton removal reaction, in addition
to the measurement of four new $\gamma$-ray transitions in $^{53}$Sc \cite{mcd10}. Although the new transitions
could not be placed in the $^{53}$Sc level scheme, the authors of Ref.~\cite{mcd10} attribute the positive-parity
states populated by the reaction to the removal of $sd$-shell protons, which highlights the role of cross-shell
excitations in such reactions. It is noted that this mechanism is important for the interpretation of $^{55}$Sc
in the present work.

The low-lying structure of the even-$A$ isotopes $^{54}$Sc and $^{56}$Sc was also reported by Crawford {\it et al.}~\cite{cra10}
from the decays of isomeric states and, in the case of $^{54}$Sc, from the $\beta$ decay of $^{54}$Ca. A $\gamma$-ray
peak at 247 keV was reported from the $^{54}$Ca decay study, which confirms the transition previously reported by
Mantica {\it et al.}~\cite{man08}. A 110-keV isomeric state in $^{54}$Sc was originally reported by Grzywacz
{\it et al.}~\cite{grz98}, and later confirmed by Refs.~\cite{lid04-2,cra10,kam12}. In the case of $^{56}$Sc,
two $\beta$-decaying states were reported \cite{lid04-2} with half-lives of 35(5) and 60(7) ms, and spin-parity
values of $(1^{+})$ and $(6^{+}, 7^{+})$, respectively, although the energies of the states could not be deduced
in that study. The half-lives of the two states were confirmed in Ref.~\cite{cra10}, where the respective values
were reported as 26(6) and 75(6) ms, and the lower-spin $\beta$-decaying state was assigned as the $^{56}$Sc ground
state. Moreover, the spins and parities of the states were reexamined, and values of $(5,6)^{+}$ were assigned to
the higher-spin isomer \cite{cra10}. The low-lying structure of $^{56}$Sc was investigated via population of a
290(30)-ns, $(4)^{+}$ isomeric level at 775 keV, and excited states at 587 and 727 keV were reported in a level
scheme that was constructed using $\gamma\gamma$ coincidence relationships \cite{cra10}; it is noted that some
of the $\gamma$ rays measured from the decay of the 290-ns isomer were first reported in Ref.~\cite{lid04-2},
although the transitions could not be placed in a level scheme in that study.

The one-neutron removal reaction was studied at relativistic energies ($\approx$420 MeV/u) for neutron-rich
Sc isotopes in Ref.~\cite{sch12}, where inclusive longitudinal momentum distributions and cross-sections are
reported for projectiles from $^{51}$Sc to $^{55}$Sc. The contributions from $\ell=1$ and $\ell=3$ orbitals
(neutron removal from the $\nu p_{3/2}$--$\nu p_{1/2}$ and $\nu f_{7/2}$--$\nu f_{5/2}$ spin-orbit partners,
respectively) were estimated by fitting experimental data with theoretical, weighted momentum distributions.
In the case of the $^{9}$Be($^{55}$Sc,$^{54}$Sc)$X$ reaction, it was deduced that the $\ell=1$ component
dominates the inclusive cross-section, with only a small contribution from the $\ell=3$ orbitals. The
negligible contribution from the $\nu f_{7/2}$ orbital was attributed to the fact that the majority of
the spectroscopic strength is located in states that lie above the neutron threshold in the residual
nucleus, $^{54}$Sc. The results also suggest that the $\nu f_{5/2}$ orbital does not play a significant
role in the one neutron-removal reaction, at least not at $N\leq34$.

While properties of the nuclear ground state have been reported for $^{55}$Sc \cite{lid04-2,man08,sor98},
where the most recent study \cite{cra10} indicates a half-life and tentative spin-parity quantum numbers
of 96(2) ms and $7/2^{-}$, respectively, no information on excited states of the $N=34$ isotope was
reported prior to the present work. It is noted that preliminary results on $^{55}$Sc are provided
in Refs.~\cite{ste15-2,ste17}. In the present article, the low-lying structure of $^{55}$Sc has been
investigated using the $^{9}$Be($^{56}$Ti,$^{55}$Sc+$\gamma$)$X$ one-proton removal and
$^{9}$Be($^{55}$Sc,$^{55}$Sc+$\gamma$)$X$ inelastic-scattering reactions in order to track directly
the development of the $N=34$ subshell closure approaching $Z=20$ and, moreover, to provide a deeper
understanding of the evolution of nuclear single-particle orbitals in systems far from the valley
of $\beta$ stability.

\section{EXPERIMENT}
% NOTE - Added Sept 2017, for second submission
The experiment was performed at the Radioactive Isotope Beam Factory, operated by RIKEN Nishina Center
and Center for Nuclear Study, University of Tokyo, using a primary beam of $^{70}$Zn$^{30+}$ ions at 345
MeV/nucleon.
% Back to original text
The BigRIPS separator \cite{kub12} was employed to produce a secondary, radioactive beam that
was optimized for the transmission of $^{55}$Sc, although $^{56}$Ti also fell within the acceptance of the
spectrometer. The secondary beam was focused on a 10-mm-thick $^{9}$Be target at the eighth focal plane
along the beam line, which was surrounded by the DALI2 $\gamma$-ray detector array \cite{tak14} to measure
photons emitted from nuclear excited states. Further downstream, the reaction products were identified using
the ZeroDegree spectrometer \cite{kub12} operating in the large-acceptance mode. Other results from the present
experiment are reported in Refs.~\cite{ste13-1,ste13-2,ste15-1,ste15-2,ste17}, where particle-identification
plots and further details on the experimental conditions are provided.
% NOTE - Added Sept 2017, for second submission
It is noted that the experimental conditions were not appropriate for determination of intrinsic angular
momenta using nucleon-removal reactions in the present work.
% Back to original text

\section{RESULTS}
\begin{figure}[!t]
\includegraphics[width=\columnwidth]{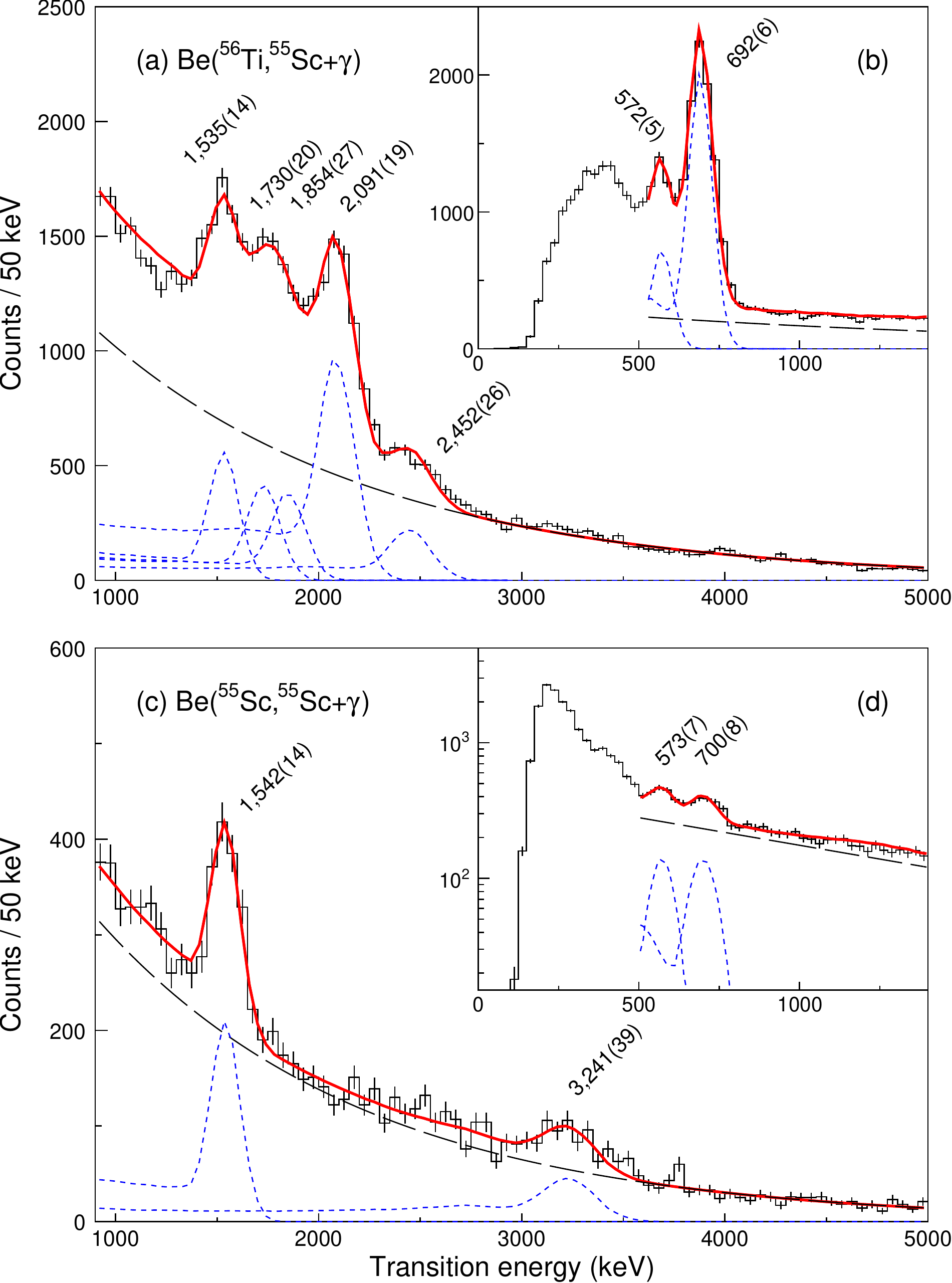}
\caption{\label{fig1}(Color online) Doppler-corrected $\gamma$-ray energy spectra for the (a)
  $^{9}$Be($^{56}$Ti,$^{55}$Sc+$\gamma$)$X$ and (c) $^{9}$Be($^{55}$Sc,$^{55}$Sc+$\gamma$)$X$
  reactions. The (black) long- and (blue) short-dashed lines are exponential fits to background
  regions and GEANT4 \cite{ago03} simulated $\gamma$-ray response functions, respectively, and
  the (red) solid lines are the total fits. Insets (b) and (d) present data from the same respective
  reactions as panels (a) and (c), but for histograms with 25 keV/bin and more restrictive angular
  cuts on DALI2 detectors (see text for details); simulations for the higher-energy peaks are not
  displayed in the insets to avoid clutter in the spectra, but are included in the total fits.
  Data with $\gamma$-ray multiplicity selections of $M_{\gamma}=1$ were used for all panels except
  for Inset (d), which presents $M_{\gamma}\geq 1$ data. Peaks are labeled by their energies in keV.}
\end{figure}

The Doppler-corrected $\gamma$-ray energy spectra deduced from the $^{9}$Be($^{56}$Ti,$^{55}$Sc+$\gamma$)$X$
and $^{9}$Be($^{55}$Sc,$^{55}$Sc+$\gamma$)$X$ reactions are presented in Figs.~\ref{fig1}(a) and \ref{fig1}(c),
respectively. It is noted that Insets \ref{fig1}(b) and \ref{fig1}(d), which display zoomed regions around the
two lowest-energy peaks reported in the present work, were deduced using a more restrictive angular cut on the
detectors of the DALI2 array in order to minimize contamination from low-energy atomic background and the
e$^{+}$e$^{-}$ annihilation peak, which lies at 511 keV in the laboratory frame of reference (more specifically,
an angular selection of $\theta\sim 52^{\circ}$--$60^{\circ}$, where $\theta$ is the polar angle relative to the
beam line, was adopted for Insets \ref{fig1}(b) and \ref{fig1}(d), whereas the main panels, Figs.~\ref{fig1}(a)
and \ref{fig1}(c), present data from detectors in the angular range $\theta\sim 52^{\circ}$--$131^{\circ}$). The
transitions, which are summarized in Table~\ref{tab1}, are reported in the present work for the first time. The
peaks at 572(4), 695(5), and 1539(10) keV were measured in both reactions, and the two energy values deduced for
each transition are consistent within uncertainties. Moreover, the peaks at 1730(20), 1854(27), 2091(19), and
2452(26) keV were only observed in the one-proton removal reaction of Fig.~\ref{fig1}(a), and the peak at 3241(39)
keV only appears in the inelastic-scattering spectrum of Fig.~\ref{fig1}(c), at least within the limit of sensitivity
of the present experiment. Errors on $\gamma$-ray energies are statistical and systematic uncertainties combined
in quadrature; the systematic component contains contributions from the energy calibration and possible shifts
in peak positions owing to indirect feeding from higher-lying states, which were estimated using the code GEANT4
\cite{ago03} by assuming excited-state lifetimes comparable to projectile times-of-flight through the reaction
target.

\begin{figure}[!t]
\includegraphics[width=\columnwidth]{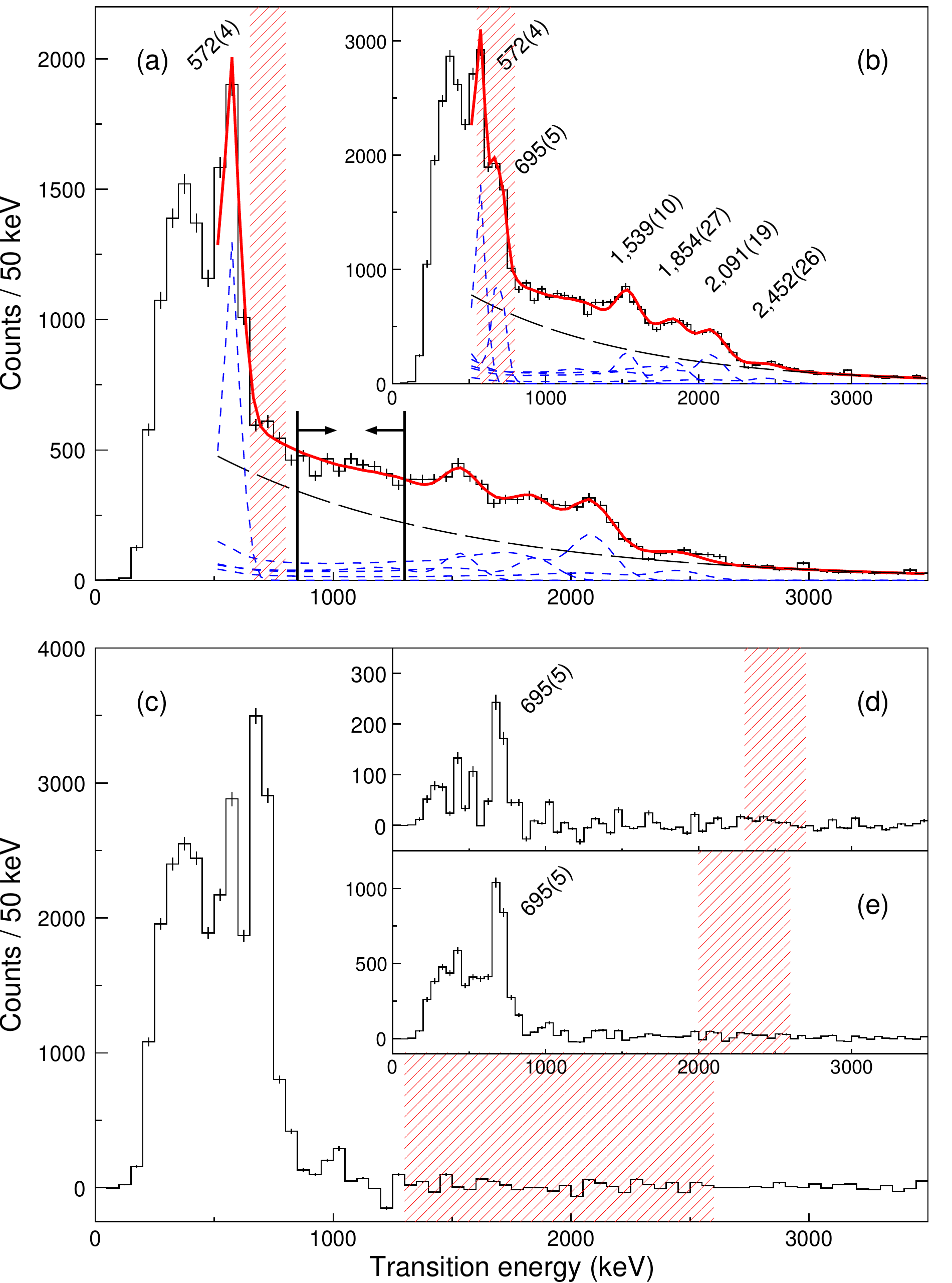}
\caption{\label{fig2}(Color online) Background-subtracted $\gamma\gamma$ coincidence relationships for transitions
  measured in the $^{9}$Be($^{56}$Ti,$^{55}$Sc+$\gamma$)$X$ reaction. The (black) long- and (blue) short-dashed lines
  in panels (a) and (b) are exponential fits to background regions and GEANT4 \cite{ago03} simulated $\gamma$-ray
  response functions, respectively, and the (red) solid lines are the total fits. The shaded regions represent
  the widths of the $\gamma$-ray energy gates applied in each panel, and the region selected for the background
  subtraction process in panel (a) is indicated by the horizontal arrows as an example. Peaks are labeled by
  their energies in keV, where given (note that the adopted, weighted-mean energies are displayed here, where
  relevant). See text for further details.}
\end{figure}

In order to place the transitions in a level scheme, $\gamma\gamma$ coincidence relationships were investigated,
which are displayed in Fig.~\ref{fig2} for the $^{9}$Be($^{56}$Ti,$^{55}$Sc+$\gamma$)$X$ reaction. It is noted that
the spectra presented in all of the panels have been background subtracted by applying $\gamma\gamma$ coincidence
gates in background regions at energies higher than the $\gamma$-ray peak values. For example, the spectrum presented
in Fig.~\ref{fig2}(a), which displays a background-subtracted $\gamma\gamma$ coincidence spectrum for the 695-keV
transition, was obtained by subtracting the normalized $\gamma\gamma$ coincidence spectrum deduced from an energy
gate set in the region between the 695- and 1539-keV peaks; the normalization factor was deduced from the total
number of events within the limits of the energy gate ($M_{\gamma}\geq1$) set in the background region, and the
number of background events within the energy gate set on the peak itself, which was estimated using fits of
the experimental $M_{\gamma}\geq1$ spectrum with simulated $\gamma$-ray response functions from the code GEANT4
\cite{ago03}. It is noted that the spectrum displayed in Fig.~\ref{fig2}(a) can be fit in a satisfactory manner
using simulated response functions for the peaks at 572, 1539, 1854, 2091, and 2452 keV, indicating coincidence
relationships between each of those five transitions and the 695-keV peak. In fact, out of all of the peaks
identified in Figs.~\ref{fig1}(a) and \ref{fig1}(b)---with the exception of the 695-keV peak itself---only
the 1730-keV transition provides no evidence for $\gamma\gamma$ coincidence relationships with the 695-keV
transition.

\begin{table}[!t]
  \caption{\label{tab1}Summary of the $\gamma$-ray transitions reported in the present work.
    Adopted (weighted-mean) values are provided for the peaks measured in both reactions. All
    energies (values listed in the first, second, and third columns) are given in keV, and the
    $\gamma$-ray relative intensities ($I_{\gamma}$) were extracted from $^{9}$Be($^{56}$Ti,$^{55}$Sc+$\gamma$)$X$
    ($M_{\gamma}\geq1$) data fitted with GEANT4 \cite{ago03} simulations assuming isotropic angular
    distributions.}
\begin{ruledtabular}
\begin{tabular}{lccr}
    Be($^{56}$Ti,$^{55}$Sc+$\gamma$)$X$ & Be($^{55}$Sc,$^{55}$Sc+$\gamma$)$X$ & Adopted   & $I_{\gamma}$ \\
    \hline
    572(5)                          & 573(7)                          & 572(4)    & 50.3(57)     \\
    692(6)                          & 700(8)                          & 695(5)    & 100(11)      \\
    1535(14)                        & 1542(14)                        & 1539(10)  & 16.8(22)     \\
    1730(20)                        & --                              & --        & 3.5(13)      \\
    1854(27)                        & --                              & --        & 14.2(21)     \\
    2091(19)                        & --                              & --        & 32.7(38)     \\
    2452(26)                        & --                              & --        & 10.0(14)     \\
    --                              & 3241(39)                        & --        & --           \\
\end{tabular}
\end{ruledtabular}
\end{table}

The spectrum of Fig.~\ref{fig2}(b) indicates the result of a wider $\gamma$-ray energy gate that encompasses
both the 572- and 695-keV peaks, and can be compared to the result of Fig.~\ref{fig2}(a) to shed light on the
$\gamma\gamma$ coincidence relationships of the 572-keV transition itself. First, it is noted that the fit of
the spectrum of Fig.~\ref{fig2}(b) requires inclusion of the simulated response function for the 695-keV peak,
which confirms the coincidence relationship between these two transitions, as discussed above. Second, it is noted
that the number of peaks required to reproduce the multiplet of transitions above 1.5 MeV remains the same; however,
the amplitude factors of the peaks in the multiplet provide further insight into which transitions lie in coincidence
with the peak at 572 keV. Indeed, one would expect the amplitude factors of the simulated response functions of the
\mbox{1539-,} \mbox{1854-,} \mbox{2091-,} and 2452-keV transitions to increase relative to the values in Fig.~\ref{fig2}(a),
because the wider energy gate applied in Fig.~\ref{fig2}(b) contains a larger number of $\gamma\gamma$ coincidence
events owing to the inclusion of the Compton component of the 695-keV transition. In fact, the increase in the number
of counts of the 695-keV transition (including counts in the full-energy photopeak and the Compton-scattered events)
within the energy gate of Fig.~\ref{fig2}(b) relative to Fig.~\ref{fig2}(a) is $\sim$1.2 and, therefore, one may
naively expect an increase in the amplitude factors of the coincident transitions of at least a similar magnitude.
Indeed, the increase of the amplitude factor for the fit of the 572-keV peak is 1.3(1), which is consistent with
the naive expectation for this transition. Similarly, the increases of the amplitude factors for the 2091- and
2452-keV $\gamma$ rays are 1.4(1) and 1.3(3), respectively, which are consistent with the value for the 572-keV
line, suggesting that no coincidence relationships exist between either of these two $\gamma$ rays and the 572-keV
transition. In the case of the peaks at 1539 and 1854 keV, however, the increases of the amplitude factors are
significantly larger---2.7(4) and 2.2(4), respectively---highlighting the coincidence relationships between each
of these two transitions and the the one at 572 keV.

Figure \ref{fig2}(c) displays the result of a $\gamma$-ray coincidence gate placed over the entire multiplet. In
this case, and similarly for Figs.~\ref{fig2}(d) and \ref{fig2}(e), which are discussed below, the coincidence gate
used for the background subtraction procedure was applied at energies higher than the 2452-keV $\gamma$ ray, which
is the highest-energy peak in the multiplet. Although the result of Fig.~\ref{fig2}(c) alone cannot be used to
distinguish which of the five transitions in the multiplet form $\gamma\gamma$ coincidences with the peaks at 572
and 695 keV, the result does, however, indicate that no transition within the multiplet forms $\gamma\gamma$ coincidence
relationships with any of the other members and, therefore, the five transitions should be placed in parallel decay paths
in the level scheme. Figs.~\ref{fig2}(d) and \ref{fig2}(e) present background-subtracted $\gamma\gamma$ coincidence
spectra for respective energy gates set on the 2452-keV peak, and a wider gate that encompasses both the 2091- and
2452-keV lines. The result of the narrow $\gamma$-ray gate---the spectrum displayed in Fig.~\ref{fig2}(d)---indicates
a coincidence relationship between the 695- and 2452-keV transitions, confirming one of the conclusions discussed
above from Fig.~\ref{fig2}(a). The result of the wider $\gamma$-ray gate---that of Fig.~\ref{fig2}(e)---indicates
$\gamma\gamma$ coincidences between the 695- and 2091-keV transitions, which is also in agreement with the result
of Fig.~\ref{fig2}(a). It is important to realize that while an increase in the number of events in the 695-keV
peak by a factor of at least $\sim$1.6 is naively expected, owing to real coincidences with the Compton component
of the 2452-keV transition that falls within the wider energy gate, the actual increase in the number of events
is notably larger than that value ($\sim$5), which highlights the additional coincidence relationship between
the 695- and 2091-keV transitions. In fact, this number is consistent with the ratio ($R$) of the number of
events of the 2452- and 2091-keV fitted response functions that lie within the respective wide and narrow
$\gamma$-ray coincidence gates of Figs.~\ref{fig2}(e) and \ref{fig2}(d), which is $R\sim5.0$ (it is noted that
a minor contribution from the 1854-keV peak within the wider energy gate was neglected, which changes $R$ by
only $\sim$0.1). Furthermore, Figs.~\ref{fig2}(d) and \ref{fig2}(e) provide no evidence for $\gamma\gamma$
coincidences between either the 2091- or the 2452-keV $\gamma$ rays and the 572-keV transition, which also
confirms the conclusions drawn from the spectra of Figs.~\ref{fig2}(a) and \ref{fig2}(b). It is noted that
statistics were insufficient to confirm the proposed $\gamma\gamma$ coincidence relationships using the
$^{9}$Be($^{55}$Sc,$^{55}$Sc+$\gamma$)$X$ inelastic-scattering data.

\section{DISCUSSION}
\begin{figure}[!t]
\includegraphics[width=\columnwidth]{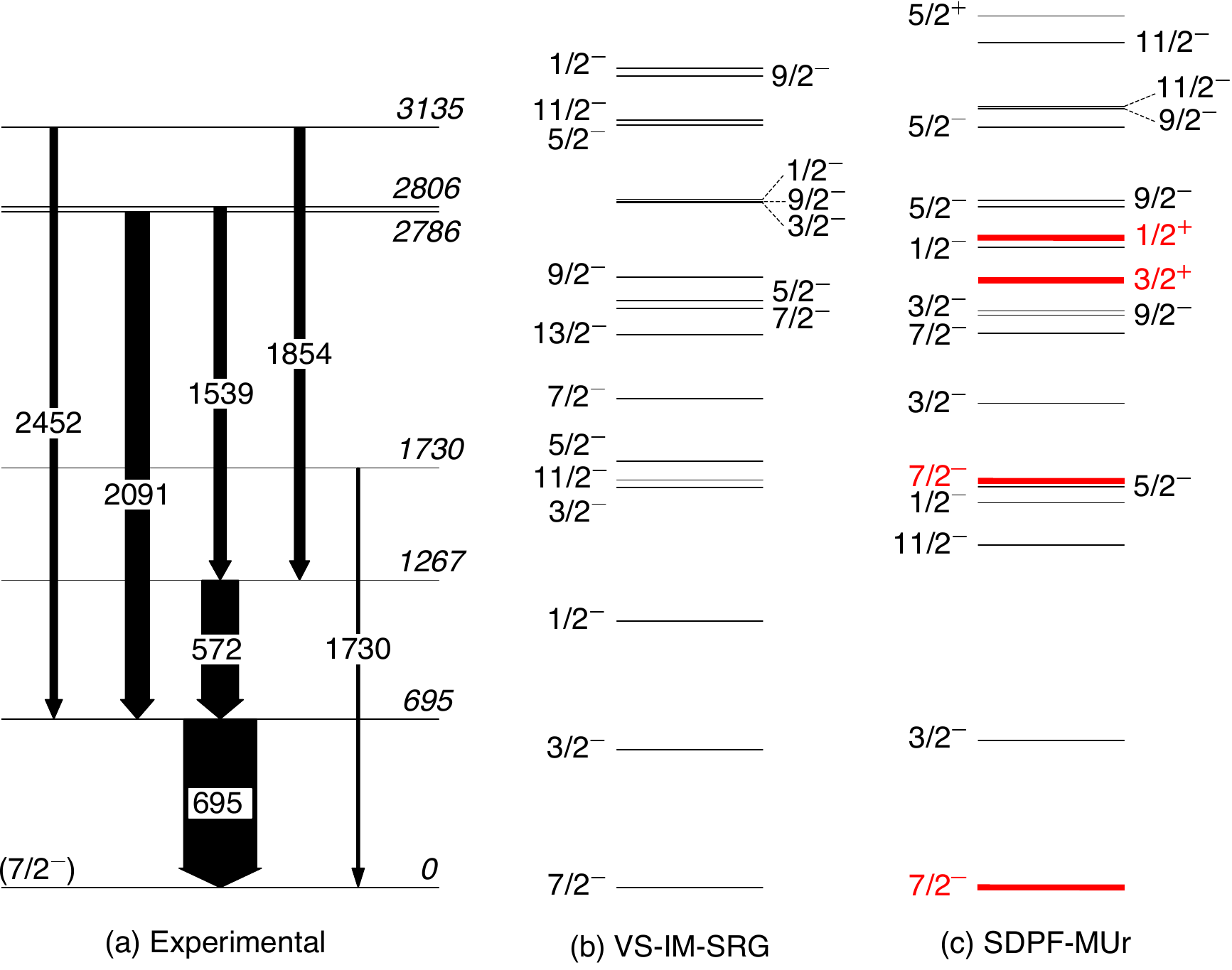}
\caption{\label{fig3}(Color online) (a) Level scheme for $^{55}$Sc deduced in the present work. The widths of the
  $\gamma$-ray lines are proportional to relative intensities measured in the $^{9}$Be($^{56}$Ti,$^{55}$Sc+$\gamma$)$X$
  reaction. The columns labeled (b) VS-IM-SRG and (c) SDPF-MUr are predictions of $^{55}$Sc spectra using the {\it ab initio}
  many-body method and large-scale shell-model calculations, respectively (see text for details). Note that a maximum of three
  states are displayed for each spin-parity in columns (b) and (c) in order to avoid clutter in the figure;
  % NOTE - Added Sept 2017, for second submission
  in column (c), the four states with sizable cross sections as predicted by nuclear reaction theory for the one-proton
  removal reaction in Table~\ref{tab2} (the $7/2^{-}$ ground state and the $7/2^{-}$, $3/2^{+}$, and $1/2^{+}$ excited states
  predicted at 1.7, 2.5, and 2.7 MeV, respectively), which is discussed in the text below, are highlighted by thick red lines
  and red text.
  % Back to original text:
  Spin-parity and energy labels on the levels are given by regular and italic fonts, respectively.}
\end{figure}

The level scheme constructed from the $\gamma\gamma$ coincidence measurements discussed above, the
$\gamma$-ray relative intensities listed in Table~\ref{tab1}, and $\gamma$-ray energy sum rules is displayed
in Fig.~\ref{fig3}(a). The 695-keV transition is placed in the level scheme feeding the ground state because
it carries the largest relative intensity and, as discussed above, forms $\gamma\gamma$ coincidence relationships
with each of the \mbox{572-,} \mbox{1539-,} \mbox{1854-,} \mbox{2091-,} and 2452-keV $\gamma$ rays. Owing to the
fact that the 2091- and 2452-keV transitions do not exhibit $\gamma\gamma$ coincidence relationships with the
peak at 572 keV, they are placed in the level scheme in parallel to that transition, feeding the 695-keV level.
Since it was deduced that the 1539- and 1854-keV $\gamma$ rays exhibit $\gamma\gamma$ coincidence relationships
with the 572-keV line, they are both placed in the level scheme feeding the state at 1267 keV. It is noted that
the energy sum of the 572(4)- and 1854(27)-keV $\gamma$ rays is consistent, within uncertainties, with the energy
of the 2452(26)-keV peak and, therefore, the 1854- and 2452-keV transitions are placed depopulating a common energy
level at 3135 keV. It is also important to realize that, although the energy sum of the 572(4)- and 1539(10)-keV
transitions is consistent with the energy of the 2091(19)-keV line, the 1539- and 2091-keV transitions are {\it not}
placed in the level scheme depopulating a common state. This is owing to the fact that while the 1539-keV line is
observed in both the $^{9}$Be($^{56}$Ti,$^{55}$Sc+$\gamma$)$X$ and $^{9}$Be($^{55}$Sc,$^{55}$Sc+$\gamma$)$X$ reactions
of Figs.~\ref{fig1}(a) and \ref{fig1}(c), respectively, the 2091-keV transition is {\it only} present in the
one-proton removal reaction, indicating that these two $\gamma$ rays must depopulate two distinct energy levels;
indeed, it is perhaps not surprising that these excited states are separated by only $\sim$20 keV considering the
density of predicted levels around 2.8 MeV displayed in Figs.~\ref{fig3}(b) and \ref{fig3}(c) (the details of these
calculations are discussed below). It was established that the 1730-keV transition does not form $\gamma\gamma$
coincidences with any of the other measured $\gamma$ rays and, therefore, it is placed in the level scheme feeding
the ground state directly. Owing to insufficient statistics in the $^{9}$Be($^{55}$Sc,$^{55}$Sc+$\gamma$)$X$ reaction,
the 3241-keV transition could not be placed in the level scheme. It should also be realized that the present experiment
was not sensitive to low-energy $\gamma$ rays ($E_{\gamma}\lesssim0.5$ MeV) owing to DALI2 detector threshold settings;
for example, $\gamma$-ray transitions between the 3135-keV level and the states at 2786 or 2806 keV ($\Delta E\lesssim350$
keV) cannot be ruled out.

We now calculate the spectrum of $^{55}$Sc using the VS-IM-SRG approach, beginning from the 1.8/2.0 (EM) NN+3N
chiral Hamiltonian developed in Refs.~\cite{heb11,sim16}. While fit to reproduce only two-, three-, and four-body
data, this interaction predicts saturation properties in infinite nuclear matter and has been shown to reproduce
ground-state energies throughout the light and medium-mass region \cite{sim17}. With all calculation details
given in Ref.~\cite{sim17}, we use the Magnus formulation of the IM-SRG \cite{mor15} to sequentially decouple
the $^{40}$Ca core as well as a $pf$-shell valence-space Hamiltonian in which 3NFs among the 15 valence nucleons
are captured via ensemble normal ordering \cite{str17}. Finally, we diagonalize with the NuShellX shell-model
code \cite{bro14} to obtain negative-parity states in $^{55}$Sc.

The resulting theoretical energy levels of $^{55}$Sc using the VS-IM-SRG and the SDPF-MUr shell-model effective
interaction (a modified version of SDPF-MU \cite{uts12-1} that includes the changes described in Ref.~\cite{ste15-1}),
which predicts positive-parity states from proton $sd$-$pf$ cross-shell excitations, are displayed in Figs.~\ref{fig3}(b)
and \ref{fig3}(c), respectively. It is noted that the tentative spin-parity assignment ($J^{\pi}$) for the ground state,
$7/2^{-}$ \cite{cra10}, is reproduced successfully by both sets of calculations. Moreover, both theories predict that
the $3/2^{-}_{1}$ level is the first excited state, and the energies of the predicted states are in good agreement
($\lesssim150$ keV) with the experimental level at 695(5) keV. It is, therefore, probable that the 695-keV level
is the first $J^{\pi}=3/2^{-}$ state. At higher energies, discrepancies between the two theories arise; for example,
VS-IM-SRG predicts the second excited state to be the $1/2^{-}_{1}$ level at $\sim$1.1 MeV, while the SDPF-MUr
Hamiltonian predicts the $1/2^{-}_{1}$ state $\sim$0.5 MeV higher, and instead the $11/2^{-}_{1}$ level is placed
above the $3/2^{-}_{1}$ state by the effective shell-model interaction. However, it should be realized that the
energy difference between the $1/2^{-}_{1}$ and $11/2^{-}_{1}$ states predicted by the SDPF-MUr Hamiltonian is not
significant ($<200$ keV). A spin-parity assignment of $J^{\pi}=1/2^{-}$ for the level at 1267 keV is likely,
because no direct decay to the $7/2^{-}$ ground state was measured; although an assignment of $J^{\pi}=5/2^{-}$
for this state cannot be completely ruled out, transition probabilities predicted by SDPF-MUr indicate that the
$5/2^{-}_{1}\rightarrow 7/2^{-}_{1}$ and $5/2^{-}_{1}\rightarrow 3/2^{-}_{1}$ transition rates are comparable and,
therefore, an assignment of $J^{\pi}=5/2^{-}_{1}$ is not consistent with the experimental level scheme. It is,
therefore, suggested that the 1267-keV state is the experimental counterpart of the $1/2^{-}_{1}$ level, and it
is noted that the energy of this state is reproduced in a satisfactory manner ($<200$ keV) by VS-IM-SRG. The
level at 1730 keV is a candidate for the $7/2^{-}_{2}$ state. According to the SDPF-MUr effective interaction,
the decay of the $7/2^{-}_{2}$ level is dominated by the transition to the $7/2^{-}_{1}$ ground state (branching
ratio $\sim$98$\%$), which is consistent with the experimental observations. Although an assignment of
$J^{\pi}=11/2^{-}_{1}$ for the 1730-keV level cannot be completely ruled out, theoretical proton-removal
calculations (discussed below) indicate sizable feeding of the $7/2^{-}_{2}$ level in the
$^{9}$Be($^{56}$Ti,$^{55}$Sc)$X$ reaction (see Table \ref{tab2}). Thus, owing to the fact
that the 1730-keV level is not populated indirectly from higher-lying states via $\gamma$-ray
decay (at least within the sensitivity of the present experiment), the most probable spin-parity
assignment for the 1730-keV state is $7/2^{-}_{2}$.

Population of positive-parity states in the $^{9}$Be($^{56}$Ti,$^{55}$Sc+$\gamma$)$X$ reaction from $sd$-$pf$
cross-shell excitations was investigated with theoretical proton-removal reaction calculations using the eikonal
model approach \cite{han03}. The single-particle cross sections for removal from each available orbital follow
the systematic approach detailed in Section III of Ref.~\cite{gad08-1}. The geometries of the complex distorting
potentials and the real potentials that bind the removed protons are deduced from the neutron and proton densities
of $^{55}$Sc and the root-mean-squared (rms) radii of the active valence and core proton orbitals, respectively,
both given by spherical Hartree-Fock (HF) calculations \cite{bro98}. A Gaussian $^{9}$Be target density with rms
radius of 2.36 fm and a zero-range effective two-nucleon (NN) interaction were also assumed in constructing the
$^{55}$Sc-target and proton-target interactions. The Woods-Saxon proton binding potentials in this case have fixed
diffuseness (0.7 fm) and spin-orbit strength (6 MeV). The deduced radius parameters, $r_{0}$, were 1.294, 1.328,
1.221, and 1.252 fm for the $0f_{7/2}$, $0f_{5/2}$, $1p_{3/2}$, and $1p_{1/2}$ valence orbitals and 1.315, 1.326, and
1.318 fm for the $0d_{5/2}$, $0d_{3/2}$, and $1s_{1/2}$ $sd$-shell core orbitals, respectively. The depth of each
potential was adjusted to reproduce the physical separation energy for the removal reaction to the final state
of interest. The ground-state to ground-state proton separation energy was 16.52 MeV \cite{wan12}, and the beam
energy at mid-target in the calculations was 200 MeV/u. The theoretical single-particle cross sections, multiplied 
by the spectroscopic factors from the nuclear structure calculations (SDPF-MUr effective interaction), predict the
partial cross sections to each final state; the theoretical spectroscopic factors and partial cross sections are
provided in Table~\ref{tab2} for reference.

\begin{table}[!t]
  \caption{\label{tab2}Theoretical spectroscopic factors ($C^{2}S$) and cross sections ($\sigma_{\scriptsize{\mbox{theory}}}$)
    for the $^{9}$Be($^{56}$Ti,$^{55}$Sc)$X$ one-proton removal reaction at 200 MeV/u for final states in $^{55}$Sc with
    energies ($E_{\scriptsize{\mbox{theory}}}$) predicted by the SDPF-MUr effective interaction. Only the states with
    $C^{2}S>0.010$ are listed here.}
\begin{ruledtabular}
\begin{tabular}{lccr}
    $E_{\scriptsize{\mbox{theory}}}$ (MeV) & $J^{\pi}$ & $C^{2}S$ & $\sigma_{\scriptsize{\mbox{theory}}}$ (mb) \\
    \hline
    0.000 & $7/2^{-}_{1}$ & 1.390 & 10.60 \\
    0.607 & $3/2^{-}_{1}$ & 0.070 &  0.55 \\
    1.676 & $7/2^{-}_{2}$ & 0.438 &  3.20 \\
    2.285 & $7/2^{-}_{3}$ & 0.028 &  0.20 \\
    2.503 & $3/2^{+}_{1}$ & 2.524 & 13.92 \\
    2.679 & $1/2^{+}_{1}$ & 1.160 &  8.03 \\
    3.594 & $5/2^{+}_{1}$ & 0.207 &  1.25 \\
    3.721 & $5/2^{+}_{2}$ & 0.049 &  0.29 \\
    3.937 & $3/2^{+}_{2}$ & 0.290 &  1.54 \\
    4.213 & $1/2^{+}_{2}$ & 0.305 &  2.02 \\
    4.238 & $3/2^{+}_{3}$ & 0.275 &  1.46 \\
\end{tabular}
\end{ruledtabular}
\end{table}

As indicated in Fig.~\ref{fig3}(c), the SDPF-MUr Hamiltonian predicts several positive-parity states at $E\geq2.5$
MeV: the respective $1/2^{+}_{1}$, $3/2^{+}_{1}$, and $5/2^{+}_{1}$ states at 2.7, 2.5, and 3.6 MeV. In the case of
the $^{9}$Be($^{54}$Ti,$^{53}$Sc+$\gamma$)$X$ one-proton removal reaction of Ref.~\cite{mcd10}, it was estimated
that $\gtrsim60\%$ of the reaction cross section populates excited states and, moreover, it was argued that a sizable
fraction of the spectroscopic strength to $^{53}$Sc excited states can be attributed to proton-hole ($sd$-$pf$ cross-shell)
excitations. In a similar manner, the proton-removal reaction theory for $^{55}$Sc in the present work suggests sizable
cross-sections for population of the $1/2^{+}_{1}$ and $3/2^{+}_{1}$ states from $sd$-shell proton-hole excitations; the
calculations indicate that the exclusive cross-sections for both states (8.0 and 13.9 mb, respectively) are comparable
to the value for the population of the $7/2^{-}$ ground state (10.6 mb), while direct population of individual
negative-parity states from the one-proton removal reaction are relatively low ($\leq0.2$ mb) with the exception
of the $3/2^{-}_{1}$ (0.5 mb) and $7/2^{-}_{2}$ (3.2 mb) states, the latter of which was discussed above and suggested
to correspond to the 1730-keV state. More specifically, the suggested spin-parity assignments for the 2786- and 3135-keV
states are $1/2^{+}_{1}$ and $3/2^{+}_{1}$, respectively. It is noted that the SDPF-MUr calculated $B(E1)$ matrix element
for the transition from the $1/2^{+}_{1}$ state to the $3/2^{-}_{1}$ state ($\sim$10$^{-4}$ $e^{2}$fm$^{2}$) dominates over
the predicted value to the $1/2^{-}_{1}$ state ($\sim$10$^{-6}$ $e^{2}$fm$^{2}$), which is consistent with the
experimental decay pattern of the 2786-keV level. It is also worthwhile noting that the predicted $1/2^{+}_{1}$ state
reproduces the energy of the level at 2786 keV rather well ($\Delta E\sim$100 keV). The $E1$ matrix elements describing
the decays to the $1/2^{-}_{1}$ and $3/2^{-}_{1}$ states from the predicted $3/2^{+}_{1}$ level are rather small, but
comparable to one another ($\sim$10$^{-6}$ $e^{2}$fm$^{2}$), and because the reaction theory indicates significant
population of the $3/2^{+}_{1}$ state in the one-proton removal reaction, the 3135-keV level is suggested to be the
experimental counterpart of this state, despite the relatively large discrepancy between the predicted and experimental
excitation energies ($\sim$0.6 MeV).

The level at 2806 keV was also populated in the $^{9}$Be($^{55}$Sc,$^{55}$Sc+$\gamma$)$X$ inelastic-scattering reaction,
and it is suggested to be a negative-parity state, although its spin value is uncertain; the fact that it is observed
to populate only the suggested $1/2^{-}_{1}$ level at 1267 keV indicates that the spin of the 2806-keV state is likely
limited to $J\leq5/2$. In the case of the one-proton removal reaction, it is probable that this state is fed indirectly
from the $\gamma$-ray decay of the 3135-keV level based on the predictions of the reaction theory calculations, which
do not indicate significant feeding of any negative-parity excited states except for the $7/2^{-}_{2}$ level; however,
as discussed above, measurements of $\gamma$-ray peaks at relatively low energies ($\lesssim0.5$ MeV) is ambiguous in
the present work owing to detector threshold settings. Thus, the $3135\rightarrow2806$-keV transition ($\sim$330 keV)
is suggested here, but requires confirmation from future measurements.

In Refs.~\cite{cra09,cra10} the structure of $^{53}$Sc was discussed in the context of the extreme single-particle
model by considering the coupling of the valence $\pi f_{7/2}$ proton to excited states of $^{52}$Ca. It was reported
that the $3/2^{-}_{1}$ state, which lies at 2.11 MeV in $^{53}$Sc \cite{cra09,cra10,mcd10}, as well as the tentative
$9/2^{-}$ and $11/2^{-}$ levels at 2.28 and 2.62 MeV \cite{bha09}, respectively, are part of the quintet of states
that result from the $\pi f_{7/2}$$\otimes$$^{52}$Ca$(2^{+}_{1})$ configuration; the fact that the energies of the states
are comparable to that of the first $2^{+}$ state of $^{52}$Ca (2.56 MeV \cite{huc85,gad06}) indicates the success of
the extreme single-particle model in this particular case and, in turn, highlights the robust nature of the $N=32$
subshell closure in Ca and Sc isotopes. Indeed, the $3/2^{-}_{1}$ level is expected to be the lowest-lying state of
the quintet of states, and is sensitive to the magnitude of the neutron shell gap at the Fermi surface. In the case
of $^{55}$Sc, the robustness of the $N=34$ subshell gap can be assessed in a similar manner by comparing the energy
of the $3/2^{-}_{1}$ level to the energy of the first $2^{+}$ state of $^{54}$Ca \cite{ste13-1}. The level at 695(5)
keV in $^{55}$Sc, which is the lowest-lying excited state identified in the present study, is suggested to be the
$3/2^{-}_{1}$ level and lies at an energy that is notably lower than $E(2^{+}_{1})$ of the Ca core---2.04(2) MeV---in
contrast to the situation at $N=32$. This result, therefore, suggests a breakdown of the extreme single-particle
model in this case owing to a rapid weakening of the $N=34$ subshell closure as protons are added to the $\pi f_{7/2}$
orbital, which confirms the suggestion made in Ref.~\cite{cra10}.
% NOTE - Added Sept 2017, for second submission
The nature of the first $3/2^{-}$ level was investigated using the SDPF-MUr Hamiltonian: this shell-model effective
interaction indicates that the $3/2^{-}_{1}$ state is dominated by the $\pi(f^{1}_{7/2})$--$\nu(p^{4}_{3/2}p^{1}_{1/2}f^{1}_{5/2})$
configuration, corresponding to the $\nu(p_{1/2}\rightarrow f_{5/2})$ neutron excitation, which contributes $71\%$
to the wave function. The $\pi(p^{1}_{3/2})$--$\nu(p^{4}_{3/2}p^{2}_{1/2})$ configuration, which is the result of the
$\pi(f_{7/2}\rightarrow p_{3/2})$ proton excitation, contributes only $12\%$ to the wave function, while other
configurations are less significant and have probabilities $<5\%$ each. It is also noted that the first $2^{+}$
state of $^{54}$Ca is dominated by the same neutron excitation as that predicted for the $3/2^{-}_{1}$ state of
$^{55}$Sc---the $\nu(p_{1/2}\rightarrow f_{5/2})$ excitation---which contributes $93\%$ to the wave function of the
$2^{+}_{1}$ excited state in the calcium isotone. Similarly, the single-particle occupancies extracted from the
VS-IM-SRG calculations suggest that the first $3/2^{-}$ state of $^{55}$Sc is predominantly based on the
$\nu(p_{1/2}\rightarrow f_{5/2})$ neutron excitation.
% Back to original text:
In the case of the Ti isotopes, it was
reported that no significant $N=34$ subshell gap is present \cite{din05,lid04-1} owing to the increased strength of
the attractive nucleon-nucleon interaction between the $\pi f_{7/2}$ and $\nu f_{5/2}$ orbitals \cite{ots05} at $Z=22$.
According to the SDPF-MUr effective interaction, the magnitude of the $\nu p_{1/2}$--$\nu f_{5/2}$ single-particle
energy gap in Ti isotopes is $\sim$1.8 MeV (calculated for $^{56}$Ti), where the gap is defined as the energy required
to promote a nucleon from the highest occupied orbital, $\nu p_{1/2}$, to the lowest unoccupied orbital, $\nu f_{5/2}$,
evaluated with the monopole interaction. Development of a weak $N=34$ subshell closure becomes apparent in the Sc
isotopes, where the magnitude of the shell gap is calculated to be $\sim$2.2 MeV for $^{55}$Sc, while a larger $N=34$
subshell closure is present in the Ca isotopes ($\sim$2.6 MeV for $^{54}$Ca) owing to the removal of the final proton
from the $\pi f_{7/2}$ orbital.
% NOTE - Added Sept 2017, for second submission
However, it is stressed that nuclear shell gaps---such as the $\nu p_{1/2}$--$\nu f_{5/2}$ single-particle
energy gap discussed here---are not experimental observables \cite{dug15}, and their magnitudes are dependent
on the adopted shell-model interaction and the valence space in which the theoretical framework is applied.
% Back to original text:
Further input on the development of the $N=34$ subshell gap around $Z=20$ should build on the result of the present
work by, for example, extracting neutron separation energies from mass measurements of Ca and Sc isotopes beyond
$N=34$.

\section{SUMMARY}
The low-lying structure of $^{55}$Sc has been investigated using the $^{9}$Be($^{56}$Ti,$^{55}$Sc+$\gamma$)$X$
one-proton removal and $^{9}$Be($^{55}$Sc,$^{55}$Sc+$\gamma$)$X$ inelastic-scattering reactions at $\sim$200
MeV/u at the RIKEN Radioactive Isotope Beam Factory. The level scheme, which was constructed using measurements
of $\gamma\gamma$ coincidence relationships and $\gamma$-ray relative intensities, was compared to theoretical
calculations using the {\it ab initio} many-body method with the valence-space formulation of the in-medium
similarity renormalization group (VS-IM-SRG) \cite{tsu12,str16,her16,str17}, large-scale shell-model calculations
with a modified SDPF-MU effective interaction (SDPF-MUr) \cite{ste15-1,uts12-1}, and proton-removal reaction
theory with the eikonal model approach \cite{han03}. The reaction theory calculations indicate sizable populations
of the $1/2^{+}_{1}$ and $3/2^{+}_{1}$ states in the one-proton removal reaction, and are suggested to correspond
to the experimental levels at 2786 and 3135 keV, respectively. The VS-IM-SRG and SDPF-MUr calculations both
predict a low-lying $3/2^{-}_{1}$ state, which is suggested to be the counterpart of the experimental level
at 695(5) keV. This state lies significantly lower in energy than the first $2^{+}$ state of $^{54}$Ca (2.04
MeV \cite{ste13-1}), and suggests a rapid weakening of the $N=34$ subshell gap as protons are added to the
$\pi f_{7/2}$ orbital.

\begin{acknowledgments}
We thank the RIKEN Nishina Center accelerator staff and BigRIPS team for their contributions to the experiment. 
This work is part of the CNS-RIKEN joint research project on large-scale nuclear-structure calculations. J.A.T.
acknowledges the support of the Science and Technology Facilities Council (UK) grant ST/L005743, and J.D.H.
acknowledges support from the National Research Council of Canada and NSERC. We thank K. Hebeler, J. Simonis,
and A. Schwenk for providing the 3N matrix elements used in this work and for valuable discussions. Computations
were performed with an allocation of computing resources at the J{\"u}lich Supercomputing Center (JURECA).
\end{acknowledgments}

\providecommand{\noopsort}[1]{}\providecommand{\singleletter}[1]{#1}%

\end{document}